\documentclass[letterpaper]{article}
\usepackage{nips,times}
\usepackage{helvet}
\usepackage{courier}
\usepackage{multirow}
\usepackage{graphicx}
\usepackage{epsfig}
\usepackage{subfig}
\usepackage{natbib} 
\usepackage{amsmath,txfonts}
\usepackage{chngpage} 

\begin{document}
%
\title{Exploring the Relationship between Membership Turnover and Productivity in Online Communities}
\author{Xiangju Qin{$^{\sharp}$}, Michael Salter-Townshend{$^{\diamondsuit}$}, P\'{a}draig Cunningham{$^{\sharp}$}\\
{$^{\sharp}$}School of Computer Science \& Informatics, University College Dublin\\
{$^{\diamondsuit}$}Department of Statistics, University of Oxford
}

\maketitle
\begin{abstract}
\begin{quote}

One of the more disruptive reforms associated with the modern Internet is the emergence of online communities working together on knowledge artefacts such as Wikipedia and OpenStreetMap. Recently it has become clear that these initiatives are vulnerable because of problems with membership turnover. This study presents a longitudinal analysis of 891 WikiProjects where we model the impact of member turnover and social capital losses on project productivity. By examining social capital losses we attempt to provide a more nuanced analysis of member turnover. In this context social capital is modelled from a social network perspective where the loss of more central members has more impact. We find that only a small proportion of WikiProjects are in a relatively healthy state with low levels of membership turnover and social capital losses. The results show that the relationship between social capital losses and project performance is U-shaped, and that member withdrawal has significant negative effect on project outcomes. The results also support the mediation of turnover rate and network density on the curvilinear relationship. 

\end{quote}
\end{abstract}

\section{Introduction}
\begin{figure}[!htb]
\baselineskip=12pt
{\fontsize{10.0pt}{10.0pt}\selectfont
\begin{center}
  \subfloat{\includegraphics[width=0.98\textwidth]{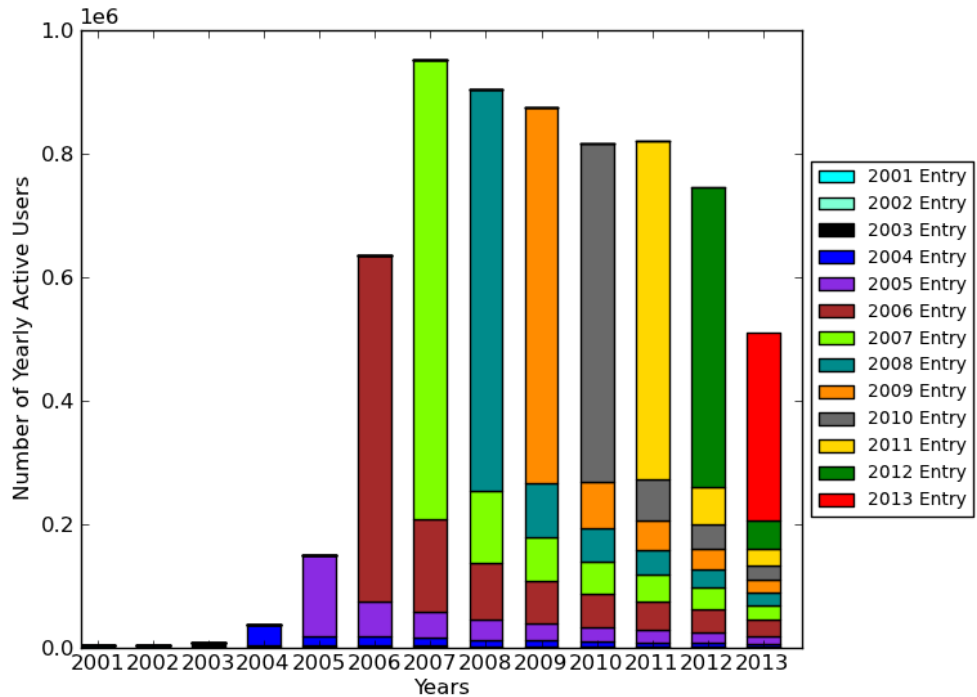}}\\
\caption{Statistics of Yearly Active Editors in English Wikipedia from 2001 to 2013. The data for 2013 was only till August 5, 2013.}
\label{fig:yearly_active_editors}
\end{center}
}
\end{figure}

With the popularity of Web 2.0, recent years have witnessed a growing population of online communities which rely on contributions from online volunteers to build knowledge and software artifacts. Despite the success of a few communities, such as Linux, Apache, Wikipedia and OpenStreetMap, many of them fail to generate desired outcomes. Many of these failures are due to the intrinsic characterstics of open collaboration in online communities, namely, high levels of member withdrawal \citep{Faraj2011,Ransbotham2011}. Even in successful online communities, such as the English version of Wikipedia, high churn in membership is still the norm, with 60\% of registered users staying only a day \citep{Panciera2009,Dabbish2012}. Figure \ref{fig:yearly_active_editors} presents the statistics of yearly active registered editors on English Wikipedia from its inception. Figure \ref{fig:yearly_active_editors} shows that each year the number of new editors dominates the amount of active editors, suggesting that Wikipedia experiences high level of membership turnover.

In traditional organizations, the cost of leaving a job can be extremely high in terms of sacrificing material or psychological benefits, such as giving up colleagues, interesting projects, corporate stock and pension benefits \citep{Mitchell2001}. Different from traditional organizations, in online communities, without committing to any tasks, projects, or conversations, the barriers to entry and exit are very low, each participant is free to join and leave \citep{Ransbotham2011}. Joining an online group can be as easy as filling in a form or writing a profile \citep{Dabbish2012}, leaving an online group is even easier by stopping contributions or showing no activities in the community. For online volunteers, leaving an online community generally incurs little cost (except losing opportunity to learn from peers). While for an online community, member withdrawal can have a significant influence on its functionality: membership turnover may improve group performance by bringing in novel skills and expertise, but it may also threaten the cognitive structures and processes that are so useful for expertise coordination \citep{Lewis2007,Faraj2011}, and inevitably disrupt routines and established social ties and further reduce its social capital \citep{Shaw2005}.

High membership turnover in online communities has attracted much attention from researchers to study the influence of turnover on community health. The most dominant view suggests that membership turnover is disadvantageous to effective collaboration, and the ability to engage and retain members represents a key aspect of success (e.g.,\cite{Arguello2006,Butler2001}). Those who leave the community, take with them not only the unique knowledge and insight, but also the experience and social relationships they have gained and established through participation. Therefore, these studies suggest that, those departures diminish the resources available to the community and may threaten its very sustainability. On the other hand, some research suggests that the unique capabilities of Web 2.0 techniques could help mitigate the negative effects of turnover (e.g.,\cite{Kane2007}). For example, platforms based on Web 2.0 techniques are typically equipped with functions to preserve all previous contributions by past members in an organized and searchable manner, so as to facilitate the utilization of past contributions for future community members as needed \citep{Kane2009,Wagner2006}. Different from previous studies, \cite{Ransbotham2011} suggested that moderate levels of turnover are best for collaboration in online communities, such that the influx of novel knowledge exceeds the loss of existing knowledge held by departing members. While these studies shed light on the relationship between turnover and performance in the online scenario, a major limitation is that they fail to consider social capital loss through member turnover and its impact on the functionality of communities.

Despite the success as a means for knowledge sharing and collaboration, little is known about the mechanism behind Web 2.0 platforms from the perspective of social network analysis. The social network approach provides a unique perspective to study the evolution of online communities over time, and to study the resilience of social networks against node removal (i.e., member drop-out) in these platforms. In this study, we take a social network approach to investigate the relationship between member turnover, social capital losses and group performance in WikiProjects. We are interested in the following questions: Does the drop-out of members in different network positions have the same effect on group performance? How does social capital loss due to departing members influence group performance? Is there consistency across the evolution of online groups? Overall, the eventual goal of this study is the provision of multi-dimensional indicators related to the health and performance of online communities, which provides insight into the management and evaluation of these communities.

The rest of this paper is organized as follows. The next section provides a brief review of related work. In Section 3, we explain network measures and project characteristics related variables, and develop hypotheses. Next, we discuss data collection and model specification. Section 5 presents the results, followed by discussion and conclusion.

\section{Membership Turnover and Performance}
There has been substantial work studying the effect of turnover on performance in traditional organizations, but perspectives on this relationship vary widely. The most dominant view suggests that turnover has a negative impact on performance, because the turnover of members can incur replacement costs, disrupt the social network, or weaken the knowledge resources of the organization \citep{Huselid1995}. A second argument indicates that in certain situations turnover may benefit organizations since departing employees are often those most dissatisfied with the current organization, so those who remain behind enjoy better working conditions and performance \citep{Krackhardt1985}. A third perspective suggests that moderate levels of turnover lead to the best organizational performance \citep{Abelson1984} in terms of creating opportunities for organizations to access new skills and knowledge through the influx of new employees \citep{Madsen2003}. In their empirical study on 38 locations of a restaurant chain, \cite{Shaw2005} found that employee turnover rate moderates the curvilinear relationship between social capital losses and performance. Despite the fact that online communities (e.g., Wikipedia) generally experience considerable membership turnover, the effects of turnover on collaborative outcomes are not well understood.

Because online communities rely on the contributions of voluntary members to survive and succeed, member turnover can have significant effects on collaborative activities in terms of resource availability and the health of a community \citep{Ransbotham2011,Wang2011}. Some studies have approached the issue from the perspective of member retention, and assumed that such retention is important to community success and survival \citep{Joyce2006}. \cite{Faraj2011} suggested that the fluidity of members can bring opportunities for knowledge collaboration when managed appropriately. By experiments on 2,065 featured articles from Wikipedia, \cite{Ransbotham2011} found a curvilinear relationship between the turnover of Wikipedia editors and the quality of an article. Specifically, they discovered that a Wikipedia article edited by experienced editors is more likely to be of higher quality up to an optimal level of experience, after which the average experience of editors decreases the quality of the article (by demoting it from FA status to lower quality status). \cite{Wang2011} provided a framework to systematically examine the relationship between member turnover and the health of a community, where community health is measured as the content volume, traffic, responsiveness, interactivity of the community. However, the authors haven't provided empirical results to evaluate their hypothesis. 

More recently, \cite{Zanetti2012} studied the evolution of social structual in open source software communities along multiple dimensions, such as the cohesiveness of the communities, the distribution of responsibilities and the resilience against membership turnover in the community. They concluded that their preliminary results are a promising step towards the definition of suitable, potentially multi-dimensional resilience and risk indicators for open source software communities. However, this study has some limitations in that it did not take into account community performance indicators.

\section{Measures}
To investigate the research questions, we explore the relationship between the performance of online voluntary groups, membership turnover, and social capital loss related measures. The following introduces the dependent, independent and control measures used in this study.

\subsection{Dependent Variable}
Previous work on Wikipedia generally measures group performance as the amount of work done by members (i.e., the total number of edits to Wikipedia article pages by members). Admittedly, this is a very raw proxy of contributions by members to Wikipedia, and it generally neglects the quality of the contributions. We further calculate the edit longevity of each edit using WikiTrust software by \cite{Adler2007}, which computes the value of each contribution by combining its quality and quantity.

\subsection{Independent Variables}
\textbf{Social Capital Losses (SC Losses)}. Social capital theory has been developed to explain the value inherent in social relationships. Social capital is created when relationships facilitate reciprocal action among people \citep{Coleman1988}. Although there is no universal and precise definition for network social capital (e.g., network closure by \cite{Coleman1988} vs structural holes by \cite{Burt1992} as social capital), social capital can be broadly defined as the benefits that network members secure from their memberships in social networks or other social structures \citep{Portes1998}. These studies further contend that organizational social capital can improve performance by enhancing commitment, increasing flexibility, and promoting knowledge sharing and information diffusion that fosters intellectual capital \citep{Coleman1988,Burt1992,Shaw2005}.

If social capital at the collective level is generated when connections facilitate instrumental action among people \citep{Coleman1988}, it is also lost when the connections among people are dissolved \citep{Burt1992,Shaw2005}. These connections, which are generally referred to as structural holes by \cite{Burt1992}, play an essential role in facilitating resources exchange and knowledge management for network members. Researchers suggest that turnover among core members who occupy key network positions should be more damaging to organizational performance than withdrawal among members who occupy less central network positions \citep{Shaw2005}. We expect that in online communities, social capital losses due to member turnover should have similar impact on group performance.

As with other researchers, we used ``betweenness centrality'' by \cite{Freeman1979} to measure the extent to which members occupied structurally advantageous positions (i.e., structural hole bridges) in the collaboration network of a WikiProject. Following \cite{Shaw2005}, we calculate social capital losses as a ratio: the average betweenness scores of the leavers divided by the average betweenness scores of all project members. 

\textbf{Turnover rate}. Measured as the percentage of members who were active in previous quarter and not active in current quarter. We consider a member being active in a quarter if and only if the member made at least one edit to articles within project scope.

\textbf{Network Density}. This variable measures the proportion of existing connections in the network to the number of possible pairwise combinations of members, and takes value from zero to one, with larger values indicating increasing density. We include this measure because previous work found that in case of social capital losses due to member turnover, a dense network with experienced members and the abundance of redundant social ties would be more tolerant to network disruptions or gaps created by turnovers \citep{Shaw2005}.  

\subsection{Control Variables}
\textbf{Quarter}. This variable indexes the quarters in which project-related measures are collected: starting with quarter 0 from the time the project is created, till the last quarter of study periods. The coefficient of this variable indicates the dynamics of project performance over time.

\textbf{Project creation time (A proxy of project age)}. Measured as the number of quarters (90-day periods) from the baseline timestamp (the timestamp that the first WikiProject was created) to the creation timestamp of a project. \cite{Chen2010} found that the creation time of a project has negative influence on its performance. They explained that due to the growth of the number of projects and participants over time, projects created later may face a  more different environment than those created earlier. In this work, we also include this variable to control for the effect.

\textbf{Project scope}. Measured as the accumulated number of articles that are tagged by a project template up to the current quarter.  Most changes reflect new articles being added to the project's scope.

\textbf{Project size}. Measured as the number of project members and active participants during a quarter.

\textbf{Edit activity in project talk pages (Discussion topics)}. In online communities, members generally rely on discussion pages to communicate with others, ask for help and support, reach consensus on controversial issues, and make collective decisions regarding rules and regulations. A large amount of content posted on project-related talk pages is beneficial for the community in terms of enabling participants to arrive at a general understanding or get help for specific questions. \cite{Qin2012} found that the number of discussion topics is positive and significant related to project efficiency in Wikipedia. We calculated the amount of group communication as the total number of discussion topics recorded in the edit history of project-related talk pages in a quarter.

\textbf{Mean membership tenure (Mean Tenure)}. The average membership tenure of a WikiProject is calculated as the sum of the membership tenure of its members divided by the number of members in a quarter. 

Following \cite{Chen2010}, we also measured membership tenure as the amount of time a member has been active in Wikipedia. Specifically, our measure of membership tenure consists of two parts: membership tenure before and after joining a WikiProject. The previous tenure of a member is calculated as the number of days between the timestamp that the user made the first edit in Wikipedia and the timestamp that this user joined a WikiProject. For a specific quarter, for any month, if a user made at least one edit to any project related pages, we calculated the monthly project tenure of this user as the number of days in that month. Then for each project member, we accumulated his previous tenure and monthly project tenure up to the current quarter to obtain his overall membership tenure in Wikipedia.

\textbf{Level of controversy}. The controversy of project topics may have impact on project performance in that it takes time and effort for project members to discuss and reach consensus on controversial issues. Following \cite{Chen2010}, we also include this variable to control for these effects and to separate them from the main effects caused by social capital losses. This variable is measured as the percentage of reverted edits to all articles within project scope during a quarter, normalized by the overall percentage of reverts in Wikipedia in the same time period.

\textbf{Tenure diversity}. In online communities like Wikipedia, it is often the case that members who have been active for a long time tend to be more experienced than new users. These active members play a fundamental role in the community in terms of spreading knowledge, information and experience across the whole community. High tenure diversity indicates high variability among project members in the time they have spent committing to group tasks and in the experience they have accumulated \citep{Chen2010}. Studies about open collaboration in online communities found that members with different tenure are more likely to perform different kinds of tasks \citep{Bryant2005}, so projects with a mix of newcomers and old-timers may enjoy better task distribution. 

The coefficient of variation has been widely used in management science literature to measure tenure diversity \cite{Chen2010}, because the potential benefits and conflict due to tenure are derived from the spread of members' experience and knowledge relative to a project's mean tenure \citep{Harrison2007}. Following this convention, we measured tenure diversity using the coefficient of variation of tenure of all project members.

\subsection{Hypotheses}
To test the effect of turnover rate, we make the first hypothesis:

\textbf{Ha} \emph{There is a negative relationship between a project's turnover rate and its performance.}

Previous work has suggested curvilinear relationship between social capital losses and organizational performance \citep{Shaw2005}. Motivated by these studies, we develop the following hypothesis regarding the effect of social capital losses:
 
\textbf{Hb} \emph{Social capital losses of a project due to member turnover relates in a curvilinear fashion to its performance, strongly declining performance with moderate levels of social capital loss but increasing performance at higher levels of social capital loss.}

\citeauthor{Shaw2005} found that turnover rate mediates the curvilinear relationship between social capital losses and organizational performance. Following \cite{Shaw2005}, we develop the following hypothesis:

\textbf{Hc1} \emph{Social capital losses and turnover rate interact in predicting project performance: the curvilinear relationship between social capital losses and performance is stronger when turnover rate is low.}

\cite{Shaw2005} suggested that in case of social capital losses, a dense network with the abundance of redundant social ties would be more tolerant to network disruptions or gaps created by turnovers. However, they found no support for their predictions regarding the mediated effect of density on the relation between social capital losses and organizational performance when density is low. They further suggested that it would be beneficial for future research to explore the conditions under which social capital losses in dense network structures are damaging to organizational performance \citep{Shaw2005}. Following this suggestion, we make the following hypothesis:

\textbf{Hc2} \emph{Social capital losses and network density interact in predicting project performance: the curvilinear relationship between social capital losses and performance is more pronounced when network density is high.}

\section{Methods}
\subsection{Data Collection}
The dataset we use in this study is extracted from the August 2013 dump of English Wikipedia\footnote{http://dumps.wikimedia.org/enwiki/20130805/}, which includes the complete edit history of all pages from the inception of Wikipedia. Project members generally claim an article in its scope by inserting project template in article talk page. We parsed the link to WikiProject in article talk pages and accumulated all articles tagged by a WikiProject up to a specific quarter to estimate its project scope. We parsed the historical edits of a project's member list to identify members and their joining time for each WikiProject. WikiProjects generally provide main pages or subpages to maintain the list of project members, and any editors can join a project by adding one's username to the member list and then leave the project by removing the username from the list.

We calculated the edit longevity of each edit to articles using the WikiTrust software by \cite{Adler2007}. We then obtained the amount of work done by project members in a quarter by accumulating all the edits or the edit longevity of edits contributed by members to project scope articles. We constructed the collaboration network for each WikiProject in a quarter by considering the co-edit activities of members to project scope articles: if two members made edits to an article, and both the edit longevity of their edits were positive, then the two members established a co-authorship connection. By accumulating all the connections among members within a project, we obtained the collaboration network for the project.

We are interested in exploring how member turnover and social capital losses influence project productivity. To make the results more meaningful, we only included those WikiProjects which had at least 50 accumulated tagged articles and at least three members (the minimal size of group) in this study. As a result, we obtained a longitudinal dataset of 891 WikiProjects each with at least 1 to 35 quarterly observations, each observation recording the characteristics, network measures and outcomes of a project in a quarter.

\subsection{Model Specification}
A preliminary analysis of the data revealed that some of the independent variables were positively skewed. Following \cite{Gelman2007}, we performed a logarithmic transformation on these independent variables to make the coefficients more comparable. To explore the possible non-linear relationship between social capital losses and project performance, we included the squares of the social capital losses in the model. To validate the possible mediated effects that member turnover or network density has on the relationship between social capital losses and performance, we added the interaction of turnover rate (or density) and social capital losses in the model. There is a non-linear dependence of productivity on social capital losses. We fit a second order polynomial (using the poly function in R) and discuss the curvilinear relationship we find. We chose a second order polynomial (quadratic) over a smoothing spline to facilitate comparison with existing work (e.g. \cite{Shaw2005}). The preliminary analysis also suggested that the collinearity of the model is very high (with kappa value larger than 30), which indicates potentially harmful collinearity among variables \citep{Baayen2008}. We used z-score to scale the $\textrm{Mean~Tenure}$ variable, which reduces the collinearity of the model significantly. To ease the interpretation, we represent the log-transformed and z-score normalization variables with``Ln" and ``Sc" prefix, respectively.

Our data has a nested structure -- quarterly observations nested within projects -- which suggest Hierarchical Linear Model \citep{Bryk1992} for our analysis. HLM is an advanced form of linear model which takes into account potential autocorrelations among time periods that are nested with the same project, while allowing us to examine the main effects of independent variables on dependent variables. Meanwhile, to account for project-level unobserved differences, we include project level random effects in the model. The model specification is as follows:
\begin{equation*}
{\fontsize{8.0pt}{8.0pt}\selectfont
\begin{aligned}[c]
\textrm{LnDV}_{jt}&=\beta_{0}+\beta_{1}\textrm{Quarter}_{jt}+\beta_{2}\textrm{Project~creation}_{jt}+\beta_{3}\textrm{LnProject~scope}_{jt}
+\beta_{4}\textrm{LnProject~size}_{jt}\\
&+\beta_{5}\textrm{LnDiscussion~topics}_{jt}+\beta_{6}\textrm{ScMean~tenure}_{jt}+\beta_{7}\textrm{Level~of~controversy}_{jt}
+\beta_{8}\textrm{Tenure~diversity}_{jt} \\
&+\beta_{9}\textrm{Turnover~rate}_{jt}+\beta_{10}\textrm{Density}_{jt}+\beta_{11}\textrm{SC~Losses}_{jt}
+\beta_{12}\textrm{SC~Losses}_{jt}^2 \\
&+\beta_{13}\textrm{SC~Losses}_{jt}\times\textrm{Turnover~rate}_{jt}
+\beta_{14}\textrm{SC~Losses}_{jt}^2\times\textrm{Turnover~rate}_{jt}+{\upsilon}_{j}
\end{aligned}
}
\end{equation*}
where variables are indexed across project ($j$) and quarter ($t$); $\textrm{LnDV}_{jt}$ represents the dependent variable; ${\upsilon}_{j}$ is the random effect at the project level. We obtained our estimates using the \textsf{lmer} function in \textsf{lme4} package \citep{lme42012} for \textbf{R} software (version 2.15.3). We check for violations of the assumption of our regression analyses and find no substantive violations.

\begin{table}[!htb]
\renewcommand{\arraystretch}{1.3}
\caption{Descriptive Statistics}
\label{Tab:Descriptive_stats}
{\fontsize{8.0pt}{8.0pt}\selectfont
\begin{center}
\begin{tabular}{|l|r|r|r|r|}
\hline
Variables & Mean & Std Dev & Min & Max  \\ \hline

Edit Longevity & 12727.3 & 39887.8 & 0 & 808926.6 \\ \hline
Edit Count & 741.5 & 2388.4 & 0 & 49665.0 \\ \hline
Quarter & 12.83 & 7.66 & 1 & 35 \\ \hline
Project creation & 18.61 & 6.1 & 0 & 45.6 \\ \hline
Project scope & 1852 & 11158.2 & 1 & 439598.0 \\ \hline
Project size & 24.55 & 4.73 & 1 & 735 \\ \hline
Discussion topics & 11.84 & 3.17 & 0 & 589 \\ \hline
Mean Tenure & 1081.4 & 452.9 & 6 & 3379 \\ \hline
Level of controversy & 0.17 & 0.11 & 0 & 2.93 \\ \hline
Tenure diversity & 0.51 & 2.01 & 0 & 1.6 \\ \hline
Turnover rate & 0.4 & 0.44 & 0 & 7 \\ \hline
Density & 0.1 & 0.08 & 0 & 0.83 \\ \hline
SC Losses & 0.29 & 0.5 & 0 & 8 \\ \hline

\end{tabular}
\end{center}
}
\end{table}
\section{Results}
The descriptive statistics and correlation matrix of the study variables are presented in Table \ref{Tab:Descriptive_stats} and Table \ref{Tab:Correlation_matrix}, respectively. It is obvious from Table \ref{Tab:Descriptive_stats} that, the amount of work (i.e., Edit Longevity and Edit Count), discussion topics, project scope and size are of reasonable variation and have a heavily right skewed distribution. By including quarter and random effects in the model, we control for the changes in overall trend of activity within WikiProjects and Wikipedia over time. We observe from Table \ref{Tab:Correlation_matrix} that the correlation between the two measures of project productivity (i.e., Log2(Edit Longevity) and Log2(Edit Count)) is positive and significant (p$<$0.001), suggesting that it is reasonable to use the edit count as a measure of project productivity.

\begin{table}[!htb]
\caption{Correlation Matrix for Variables}
\begin{adjustwidth}{-1.3in}{-1.2in}
\renewcommand{\arraystretch}{1.3}
{\fontsize{8.0pt}{8.0pt}\selectfont
\label{Tab:Correlation_matrix}
\begin{center}
\begin{tabular}{lcccccccccccccc}
  \hline
 & 1 & 2 & 3 & 4 & 5 & 6 & 7 & 8 & 9 & 10 & 11 & 12 & 13 & 14 \\  
  \hline
1. Log2(Edit Longevity) & - &  &  &  &  &  &  &  &  &  &  &  &  &  \\ 
2. Log2(Edit Count) & 0.92 & - &  &  &  &  &  &  &  &  &  &  &  &  \\ 
3. Quarter & 0.04 & 0.07 & - &  &  &  &  &  &  &  &  &  &  &  \\ 
4. Project creation & -0.29 & -0.29 & -0.21 & - &  &  &  &  &  &  &  &  &  &  \\ 
5. LnProject Scope & 0.49 & 0.59 & 0.49 & -0.22 & - &  &  &  &  &  &  &  &  &  \\ 
6. LnProject size & 0.71 & 0.75 & -0.00 & -0.42 & 0.50 & - &  &  &  &  &  &  &  &  \\ 
7. LnDiscussion topics & 0.60 & 0.64 & -0.07 & -0.46 & 0.39 & 0.77 & - &  &  &  &  &  &  &  \\ 
8. ScMean Tenure & -0.01 & 0.01 & 0.57 & 0.08 & 0.31 & -0.16 & -0.18 & - &  &  &  &  &  &  \\ 
9. Level of controversy & -0.05 & -0.12 & 0.09 & -0.09 & -0.07 & -0.05 & -0.00 & 0.03 & - &  &  &  &  &  \\ 
10. Tenure Diversity & 0.26 & 0.24 & -0.15 & -0.11 & 0.03 & 0.39 & 0.22 & -0.43 & 0.00 & - &  &  &  &  \\ 

11. Turnover rate & -0.15 & -0.17 & 0.03 & 0.02 & -0.05 & -0.13 & -0.10 & -0.01 & 0.04 & 0.02 & - &  &  &  \\
12. Density & -0.06 & -0.07 & -0.22 & 0.24 & -0.19 & -0.21 & -0.18 & -0.03 & -0.09 & 0.00 & 0.22 & - &  &  \\ 
13. SC Losses & -0.08 & -0.11 & -0.02 & 0.00 & -0.07 & -0.03 & -0.05 & -0.07 & 0.04 & 0.04 & 0.23 & 0.16 & - &  \\ 
14. $\textrm{SC~Losses}^2$ & -0.05 & -0.05 & -0.00 & 0.07 & -0.03 & -0.10 & -0.06 & 0.03 & -0.03 & -0.05 & -0.17 & -0.02 & 0.00 & - \\
   \hline

\end{tabular}
\end{center}
}
\end{adjustwidth}
\end{table}

\begin{table}[!htb]
\caption{HLM results of Predicting Amount of Work Done by Project Members (Using Maximum Likelihood estimates)}
\begin{adjustwidth}{-1.3in}{-1.2in}
\label{Tab:HLM-results}
\renewcommand{\arraystretch}{1.3}
{\fontsize{8.0pt}{8.0pt}\selectfont
\begin{center}
\begin{tabular}{lr@{}lr@{}lr@{}lr@{}lr@{}lr@{}lr@{}lr@{}lr@{}lr@{}l}
\hline
\multirow{2}{*}{Variables} & \multicolumn{ 4}{c}{\textbf{Model 1}} & \multicolumn{ 4}{c}{\textbf{Model 2}} & \multicolumn{ 4}{c}{\textbf{Model 3}} & \multicolumn{ 4}{c}{\textbf{Model 4a}} & \multicolumn{ 4}{c}{\textbf{Model 4b}} \\
\cline{ 2- 21}
 & \multicolumn{2}{c}{Coeff.} & \multicolumn{2}{c}{S. E.} & \multicolumn{2}{c}{Coeff.} & \multicolumn{2}{c}{S. E.} & \multicolumn{2}{c}{Coeff.} & \multicolumn{2}{c}{S. E.} & \multicolumn{2}{c}{Coeff.} & \multicolumn{2}{c}{S. E.} & \multicolumn{2}{c}{Coeff.} & \multicolumn{2}{c}{S. E.} \\ \hline

\multicolumn{21}{l}{\multirow{2}{*}{\textbf{Log2(Edit longevity)} as dependent variable (\#Obs: 16048, \#WikiProjects: 890, \#Quarters:1--35)}} \\
\multicolumn{21}{l}{} \\
Intercept & 3&.493{$^{***}$} & 0&.259 & 4&.115{$^{***}$} & 0&.259 & 4&.121{$^{***}$} & 0&.26 & 4&.119{$^{***}$} & 0&.259 & 4&.122{$^{***}$} & 0&.259 \\
Quarter & -0&.045{$^{***}$} & 0&.004 & -0&.043{$^{***}$} & 0&.004 & -0&.043{$^{***}$} & 0&.004 & -0&.043{$^{***}$} & 0&.004 & -0&.043{$^{***}$} & 0&.004 \\
Project creation & -0&.012 & 0&.01 & -0&.03{$^{**}$} & 0&.01 & -0&.031{$^{**}$} & 0&.01 & -0&.03{$^{**}$} & 0&.01 & -0&.03{$^{**}$} & 0&.01 \\
LnProject scope & 0&.241{$^{***}$} & 0&.014 & 0&.253{$^{***}$} & 0&.014 & 0&.253{$^{***}$} & 0&.014 & 0&.251{$^{***}$} & 0&.014 & 0&.253{$^{***}$} & 0&.014 \\
LnProject size & 1&.386{$^{***}$} & 0&.031 & 1&.249{$^{***}$} & 0&.032 & 1&.248{$^{***}$} & 0&.032 & 1&.246{$^{***}$} & 0&.032 & 1&.247{$^{***}$} & 0&.032 \\
LnDiscussion topics & 0&.16{$^{***}$} & 0&.014 & 0&.159{$^{***}$} & 0&.014 & 0&.159{$^{***}$} & 0&.014 & 0&.161{$^{***}$} & 0&.014 & 0&.161{$^{***}$} & 0&.014 \\
ScMean tenure & 0&.255{$^{***}$} & 0&.033 & 0&.246{$^{***}$} & 0&.032 & 0&.245{$^{***}$} & 0&.032 & 0&.239{$^{***}$} & 0&.032 & 0&.246{$^{***}$} & 0&.032 \\
Level of controversy & -0&.566{$^{**}$} & 0&.164 & -0&.518{$^{**}$} & 0&.162 & -0&.516{$^{**}$} & 0&.162 & -0&.508{$^{**}$} & 0&.162 & -0&.508{$^{**}$} & 0&.162 \\
Tenure diversity & 0&.697{$^{***}$} & 0&.101 & 0&.74{$^{***}$} & 0&.1 & 0&.74{$^{***}$} & 0&.1 & 0&.724{$^{***}$} & 0&.1 & 0&.74{$^{***}$} & 0&.1 \\
Turnover rate &  & & &  & -0&.478{$^{***}$} & 0&.032 & -0&.458{$^{***}$} & 0&.033 & -0&.397{$^{***}$} & 0&.036 & -0&.461{$^{***}$} & 0&.033 \\
Density$^{\dagger}$ & & & & & 2&.596{$^{***}$} & 0&.19 & 2&.616{$^{***}$} & 0&.19 & 2&.495{$^{***}$} & 0&.193 & 2&.511{$^{***}$} & 0&.194 \\
SC Losses$^{\dagger}$ &  & & &  & -11&.341{$^{***}$} & 1&.622 & -11&.722{$^{***}$} & 1&.626 & -5&.838{$^{**}$} & 2&.224 & -7&.476{$^{**}$} & 2&.851 \\
$\textrm{SC~Losses}^2$$^{\dagger}$ &  & & &  &  & & &  & 5&.38{$^{**}$} & 1&.596 & 3&.834{$^{t}$} & 2&.151 & 0&.215 & 2&.827 \\
SC Losses$^{\dagger}$ $\times$ Turnover rate &  & & &  &  & & &  &  & & &  & -12&.263{$^{***}$} & 3&.288 &  & & &  \\
$\textrm{SC~Losses}^2$$^{\dagger}$ $\times$ Turnover rate &  & & &  &  & & &  &  & & &  & 2&.493 & 3&.699 &  & & &  \\
SC Losses$^{\dagger}$ $\times$ Density$^{\dagger}$ &  & & &  &  & & &  &  & & &  &  & & &  & -34&.47{$^{*}$} & 17&.283 \\
$\textrm{SC~Losses}^2$$^{\dagger}$ $\times$ Density$^{\dagger}$ &  & & &  &  & & &  &  & & &  &  & & &  & 40&.64{$^{*}$} & 17&.281 \\
\multicolumn{21}{l}{\multirow{2}{*}{Fit}} \\
\multicolumn{21}{l}{} \\
AIC &  \multicolumn{4}{l}{61235.0} &  \multicolumn{4}{l}{60862.0}  &  \multicolumn{4}{l}{60852.0}  &  \multicolumn{4}{l}{60839.0}  &  \multicolumn{4}{l}{60848.0}  \\ 
BIC &  \multicolumn{4}{l}{61320.0} &  \multicolumn{4}{l}{60969.0}  &  \multicolumn{4}{l}{60968.0}  &  \multicolumn{4}{l}{60970.0}  &  \multicolumn{4}{l}{60979.0}  \\ 
logLik &  \multicolumn{4}{l}{-30607.0} &  \multicolumn{4}{l}{-30417.0}  &  \multicolumn{4}{l}{-30411.0}  &  \multicolumn{4}{l}{-30403.0}  &  \multicolumn{4}{l}{-30407.0}  \\ 
$\chi^2$~/~df vs. previous nested model &  \multicolumn{4}{l}{} &  \multicolumn{4}{l}{379.53{$^{***}$}}  &  \multicolumn{4}{l}{11.36{$^{***}$}}  &  \multicolumn{4}{l}{17.19{$^{***}$}}  &  \multicolumn{4}{l}{8.10{$^{*}$}}  \\ \hline
\multirow{2}{*}{Variables} & \multicolumn{ 4}{c}{\textbf{Model 1}} & \multicolumn{ 4}{c}{\textbf{Model 2}} & \multicolumn{ 4}{c}{\textbf{Model 3}} & \multicolumn{ 4}{c}{\textbf{Model 4a}} & \multicolumn{ 4}{c}{\textbf{Model 4b}} \\
\cline{ 2- 21}
 & \multicolumn{2}{c}{Coeff.} & \multicolumn{2}{c}{S. E.} & \multicolumn{2}{c}{Coeff.} & \multicolumn{2}{c}{S. E.} & \multicolumn{2}{c}{Coeff.} & \multicolumn{2}{c}{S. E.} & \multicolumn{2}{c}{Coeff.} & \multicolumn{2}{c}{S. E.} & \multicolumn{2}{c}{Coeff.} & \multicolumn{2}{c}{S. E.} \\ \hline
\multicolumn{21}{l}{\multirow{2}{*}{\textbf{Log2(Edit count)} as dependent variable (\#Obs: 16171, \#WikiProjects: 891, \#Quarters: 1--35)}} \\
\multicolumn{21}{l}{} \\

Intercept & 0&.016 & 0&.18 & 0&.474{$^{**}$} & 0&.178 & 0&.48{$^{**}$} & 0&.178 & 0&.475{$^{**}$} & 0&.178 & 0&.484{$^{**}$} & 0&.178 \\
Quarter & -0&.025{$^{***}$} & 0&.003 & -0&.023{$^{***}$} & 0&.003 & -0&.023{$^{***}$} & 0&.003 & -0&.023{$^{***}$} & 0&.003 & -0&.023{$^{***}$} & 0&.003 \\
Project creation & 0&.014{$^{*}$} & 0&.007 & -0&.0001 & 0&.007 & -0&.001 & 0&.007 & -0&.0003 & 0&.007 & -0&.0003 & 0&.007 \\
LnProject scope & 0&.299{$^{***}$} & 0&.01 & 0&.307{$^{***}$} & 0&.01 & 0&.307{$^{***}$} & 0&.01 & 0&.307{$^{***}$} & 0&.01 & 0&.307{$^{***}$} & 0&.01 \\
LnProject size & 1&.132{$^{***}$} & 0&.022 & 1&.031{$^{***}$} & 0&.022 & 1&.03{$^{***}$} & 0&.022 & 1&.031{$^{***}$} & 0&.022 & 1&.03{$^{***}$} & 0&.022 \\
LnDiscussion topics & 0&.157{$^{***}$} & 0&.01 & 0&.155{$^{***}$} & 0&.01 & 0&.155{$^{***}$} & 0&.01 & 0&.155{$^{***}$} & 0&.01 & 0&.155{$^{***}$} & 0&.01 \\
ScMean tenure & 0&.068{$^{**}$} & 0&.022 & 0&.06{$^{**}$} & 0&.022 & 0&.059{$^{**}$} & 0&.022 & 0&.058{$^{**}$} & 0&.022 & 0&.059{$^{**}$} & 0&.022 \\
Level of controversy & -1&.161{$^{***}$} & 0&.114 & -1&.131{$^{***}$} & 0&.112 & -1&.131{$^{***}$} & 0&.112 & -1&.129{$^{***}$} & 0&.112 & -1&.127{$^{***}$} & 0&.112 \\
Tenure diversity & 0&.14{$^{*}$} & 0&.069 & 0&.174{$^{*}$} & 0&.068 & 0&.175{$^{*}$} & 0&.068 & 0&.168{$^{*}$} & 0&.068 & 0&.172{$^{*}$} & 0&.068 \\

Turnover rate &  & & &  & -0&.345{$^{***}$} & 0&.022 & -0&.332{$^{***}$} & 0&.022 & -0&.315{$^{***}$} & 0&.025 & -0&.333{$^{***}$} & 0&.022 \\
Density$^{\dagger}$ & & & & & 2&.146{$^{***}$} & 0&.13 & 2&.164{$^{***}$} & 0&.13 & 2&.132{$^{***}$} & 0&.132 & 2&.103{$^{***}$} & 0&.133 \\

SC Losses$^{\dagger}$ &  & & &  & -13&.541{$^{***}$} & 1&.126 & -13&.801{$^{***}$} & 1&.127 & -11&.143{$^{***}$} & 1&.547 & -13&.549{$^{***}$} & 1&.961 \\
$\textrm{SC~Losses}^2$$^{\dagger}$ &  & & &  &  & & &  & 3&.939{$^{***}$} & 1&.096 & 4&.843{$^{**}$} & 1&.671 & -0&.109 & 1&.97 \\
SC Losses$^{\dagger}$ $\times$ Turnover rate &  & & &  &  & & &  &  & & &  & -5&.709{$^{*}$} & 2&.266 &  & & &  \\
$\textrm{SC~Losses}^2$$^{\dagger}$ $\times$ Turnover rate &  & & &  &  & & &  &  & & &  & -2&.404 & 2&.823 &  & & &  \\
SC Losses$^{\dagger}$ $\times$ Density$^{\dagger}$ &  & & &  &  & & &  &  & & &  &  & & &  & -3&.978 & 11&.692 \\
$\textrm{SC~Losses}^2$$^{\dagger}$ $\times$ Density$^{\dagger}$ &  & & &  &  & & &  &  & & &  &  & & &  & 32&.119{$^{*}$} & 12&.992 \\

\multicolumn{21}{l}{\multirow{2}{*}{Fit}} \\
\multicolumn{21}{l}{} \\

AIC &  \multicolumn{4}{l}{50019.0} &  \multicolumn{4}{l}{49473.0}  &  \multicolumn{4}{l}{49462.0}  &  \multicolumn{4}{l}{49459.0}  &  \multicolumn{4}{l}{49460.0}  \\ 

BIC &  \multicolumn{4}{l}{50103.0} &  \multicolumn{4}{l}{49580.0}  &  \multicolumn{4}{l}{49577.0}  &  \multicolumn{4}{l}{49590.0}  &  \multicolumn{4}{l}{49550.0}  \\ 

logLik &  \multicolumn{4}{l}{-24998.0} &  \multicolumn{4}{l}{-24722.0}  &  \multicolumn{4}{l}{-24716}  &  \multicolumn{4}{l}{-24713.0}  &  \multicolumn{4}{l}{-24713}  \\ 

$\chi^2$~/~df vs. previous nested model &  \multicolumn{4}{l}{} &  \multicolumn{4}{l}{552.03{$^{***}$}}  &  \multicolumn{4}{l}{12.90{$^{***}$}}  &  \multicolumn{4}{l}{6.35{$^{*}$}}  &  \multicolumn{4}{l}{6.11{$^{*}$}}  \\ \hline

\multicolumn{21}{l}{Note. $\dagger$ one quarter lag for network measures.}  \\ 
\multicolumn{21}{l}{Significance level: $^{\ast\ast\ast}$ p$<$0.001, $^{\ast\ast}$ p$<$0.01, $^{\ast}$ p$<$0.05, $^{t}$ p$<$0.1.}  \\
\end{tabular}
\end{center}
}
\end{adjustwidth}
\end{table}

Table \ref{Tab:HLM-results} presents the results of our empirical analysis in an incremental manner, with the upper and lower panel including the results when the edit longevity and edit count is used as the dependent variable, respectively. Model 1 is the baseline model including all control variables, Model 2 adds the linear terms for the independent variables into Model 1, Model 3 adds the quadratic term of $\textrm{SC~Losses}$ into Model 2, Model 4a and Model 4b add the interaction terms into Model 3. Comparison of $\chi^2$ reveals that models with interaction terms (Model 4a and Model 4b) fit the data better than the simpler models.

Our results in Model 1 about the significance of control variables are consistent with the findings by \cite{Chen2010}: Five control variables have significant effects on the amount of work (edit longevity or edit count) done by project members (p$<$.001). Specifically:
\begin{itemize}
    \item A project would be more productive if it is of larger scope (0.241 and 0.299) and size (1.386 and 1.132), or if it deals with a less controversial topic (-0.566 and -1.161).
    \item The negative and significant coefficient for quarter variable (-0.045 and -0.025) suggests that when holding other variables constant, projects generally become less productive over time.
    \item Project creation time has very weak effect on edit longevity ($\beta$=-0.045, p$>$0.1) or edit count ($\beta$=0.014, p$<$0.01) if we observe quarterly productivity for projects over a longer period of time, which is the case in this study (observation over 8 years).
    \item The average member tenure and tenure diversity are positively and significantly associated with the amount of work, suggesting that projects with high level of tenure diversity enjoy better group outcomes.
\end{itemize}

In Model 1, the positive and significant coefficient for the discussion topics variable implies that when controlling for other factors, having more discussion topics in project talk pages generally makes a project more productive. Consistent with the results by \cite{Qin2012}, this finding suggests that the amount of information and knowledge resources exchanging among participants via communication in project talk pages has a positive impact on project productivity and efficiency.

In Model 2, all three independent variables have significant effects on group productivity (p$<$.001). The coefficient for network density is positive and significant (2.596 and 2.146), suggesting that when controling for other factors, increasing the connectedness of the collaboration network for a project generally improves its productivity. The negative and significant coefficient for turnover rate (-0.478 and -0.345) implies that controlling for other factors, projects with high levels of turnover in general accomplish less work in a quarter. This provides support for hypothesis \textbf{Ha}.

\begin{figure}[!htb]
\baselineskip=12pt
{\fontsize{10.0pt}{10.0pt}\selectfont
\begin{center}
  \subfloat{\includegraphics[width=0.68\textwidth]{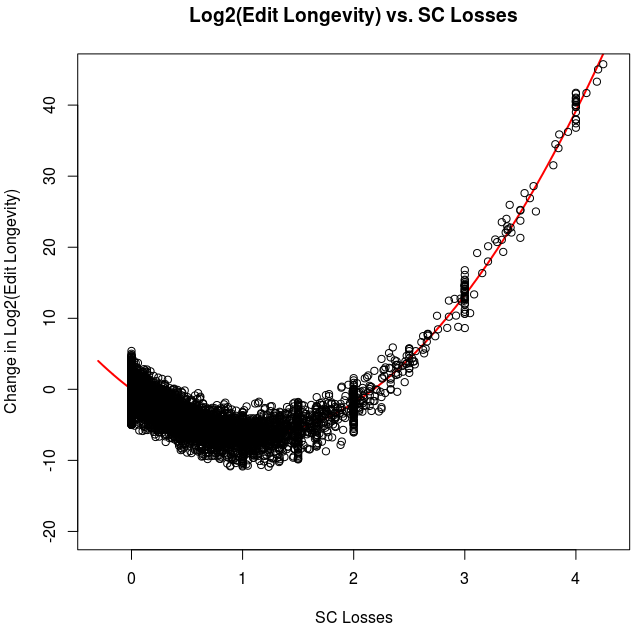}}\\
  \subfloat{\includegraphics[width=0.68\textwidth]{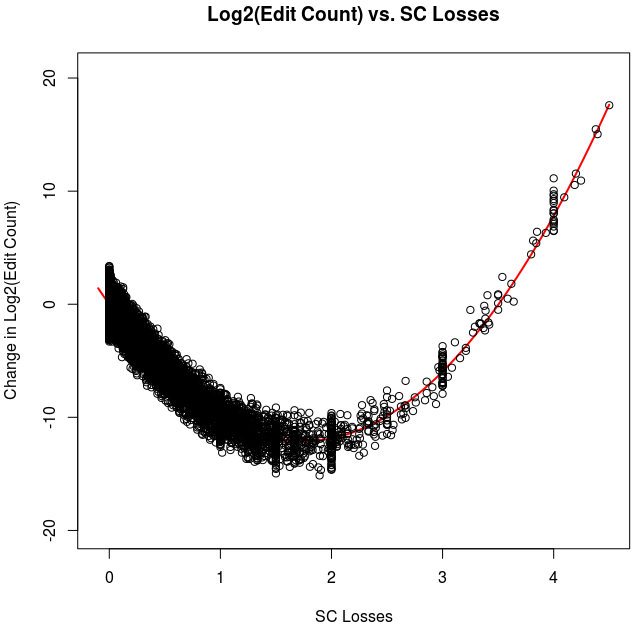}}\\
\end{center}
\caption{A U-shaped relationship for social capital losses and productivity of WikiProjects. The red lines show the model fit for the relationship of $\textrm{SC Losses}$ to $\textrm{Log2(Edit~Longevity)}$ on the left and $\textrm{Log2(Edit~Count)}$ on the right. The y-values of the dots correspond to the observed productivity minus the productivity predicted by all other variables. Thus the scatter plots depicts the covariance of productivity ( $\textrm{Log2(Edit~Longevity)}$ and $\textrm{Log2(Edit~Count)}$) with $\textrm{SC Losses}$ conditional on all other variables.}
\label{fig:sc_losses_productivity}
}
\end{figure}
Model 3 shows that both the linear and squared term of $\textrm{SC Losses}$ are significant in the expected direction. Specifically, the linear term of social capital losses has a negative main effect ($\beta$=-11.722, p$<$0.001 in the edit longevity  model and $\beta$=-13.801, p$<$0.001 in the edit count model) and its quadratic term has a positive effect ($\beta$=5.38, p$<$0.01 for edit longevity and $\beta$=3.939, p$<$0.001 for edit count). As shown in Figure \ref{fig:sc_losses_productivity}, as social capital losses increases from a very low level to a moderate level, the amount of work done by project members declines; after the moderate level, with any further increased in social capital losses from turnover, the amount of work done by project members increases. This finding is consistent with the results by \cite{Shaw2005}. One explanation for this negative curvilinear relationship between social capital losses and group productivity can be found in \cite{Shaw2005}: initially, when social capital losses are low, the loss of structural hole spanners will create the first communication gaps in a network (can also be knowledge gaps or other gaps), which will have a negative effect on group performance; when social capital losses are high, key network relationships have not been well established and maintained, which should be less damaging to group performance. The results provide full support for hypothesis \textbf{Hb}.

\begin{figure}[!htb]
\baselineskip=12pt
{\fontsize{10.0pt}{10.0pt}\selectfont
\begin{adjustwidth}{-.3in}{-.3in}
\begin{center}
  \subfloat{\includegraphics[width=0.68\textwidth]{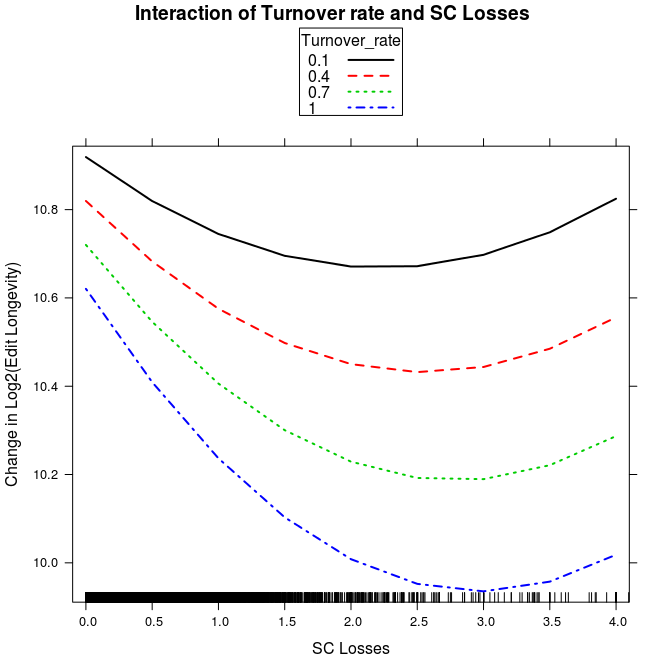}}\quad
  \subfloat{\includegraphics[width=0.68\textwidth]{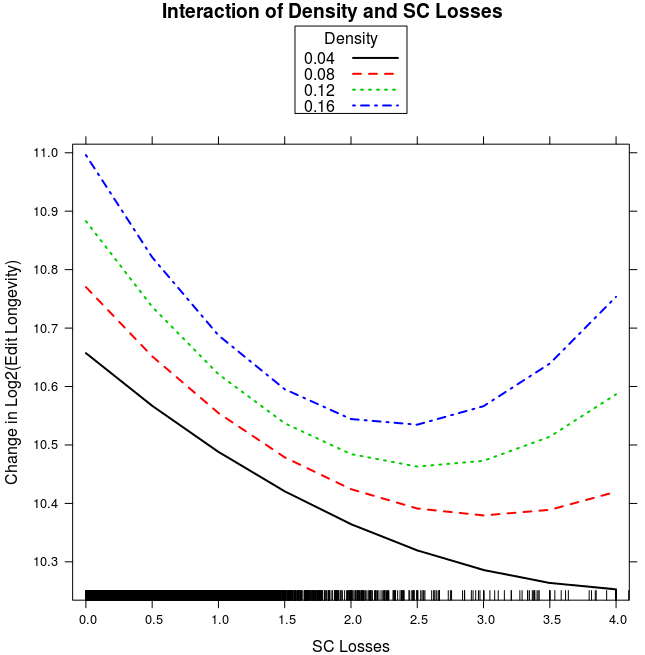}}\\
\end{center}
\end{adjustwidth}
\caption{Visualization of the mediated effects of Turnover rate and Network Density on the relationship between social capital losses and project productivity when the edit longevity is the dependent variable.The curves are generated when holding the values of turnover rate (or network density) at a specific level.}
\label{fig:Interaction_Effects}
}
\end{figure}
Hypothesis \textbf{Hc1} states that the negative curvilinear relationship between social capital losses and project productivity is more pronounced when turnover rate is low. In Model 4a, the interaction of turnover rate and the linear term of $\textrm{SC Losses}$ is significant in the expected direction, but the interaction with the squared term is insignificant. We plotted the coefficients to examine the level of support for \textbf{Hc1}. The left figure in Figure \ref{fig:Interaction_Effects} shows this relationship when the edit longevity is the dependent variable\footnote{We observe a similar trend when the edit count is the dependent variable. The figures are omitted due to space limitation.}. Consistent with \cite{Shaw2005}, we also observe that project performance is the most pronounced when social capital losses and turnover rate are at their lowest levels. The figure shows that, with low turnover rate, the curvilinear relationship is negative at low levels of social capital losses, but this effect becomes attenuated as these losses increase beyond moderate levels. One explanation for this can be explained by \cite{Shaw2005}: Members in long-standing groups generally establish routine, trust and norms of reciprocity that reduce opportunistic behaviour. Thus, the effects of the loss of structural hole bridgers should be more pronounced when turnover is low -- a situation in which reciprocity norms and routines are well-established -- than when turnover is high, so key network relationships have not been well maintained. The results provide substantial support for \textbf{Hc1}.

Hypothesis \textbf{Hc2} states that the negative curvilinear relationship between social capital losses and project productivity is more pronounced when network density is high. In Model 4b, the interaction of density and $\textrm{SC Losses}$ is significant (p$<$0.05) in the expected direction when the edit longevity is the dependent variable, but only the interaction with the squared term is significant (p$<$0.05) when the edit count is the dependent variable. The right figure in Figure \ref{fig:Interaction_Effects} visualizes this relationship. Different from the left figure in Figure \ref{fig:Interaction_Effects}, we observe in the right figure that project performance is the highest when social capital losses are at the lowest levels and network density is at higher level. One explanation for this may be about the nature of the collaboration network: a dense collaboration network implies more collaboration among members to work on group tasks, thus promoting trust, reciprocity norms and routines in the group, which has potentially positive effect on group performance. We also observe that as social capital losses increase from low to moderate levels, the slope of the line becomes steeper when density is high than that when density is low. One explanation may be: Dense networks indicate more collaborations among members, so dense networks are more susceptible to social capital losses. Moreover, the overall network density in this study is not very high as we see from Table \ref{Tab:Descriptive_stats}, there is less redundancy of social ties that would make the networks tolerant to disruptions. The figure shows that the negative curvilinear relationship is more pronounced when network density is high. Thus, \textbf{Hc2} is partially supported.

\section{Discussion}
We have found that there is a negative curvilinear relationship between social capital losses and group productivity in WikiProjects -- as social capital losses increases from very low level to moderate level, the amount of work done by project members declines; after the moderate level, with any further increased in social capital losses from turnover, the amount of work done by project members increases.

\begin{figure}[!htb]
\baselineskip=12pt
{\fontsize{10.0pt}{10.0pt}\selectfont
\begin{center}
  \subfloat{\includegraphics[width=0.90\textwidth]{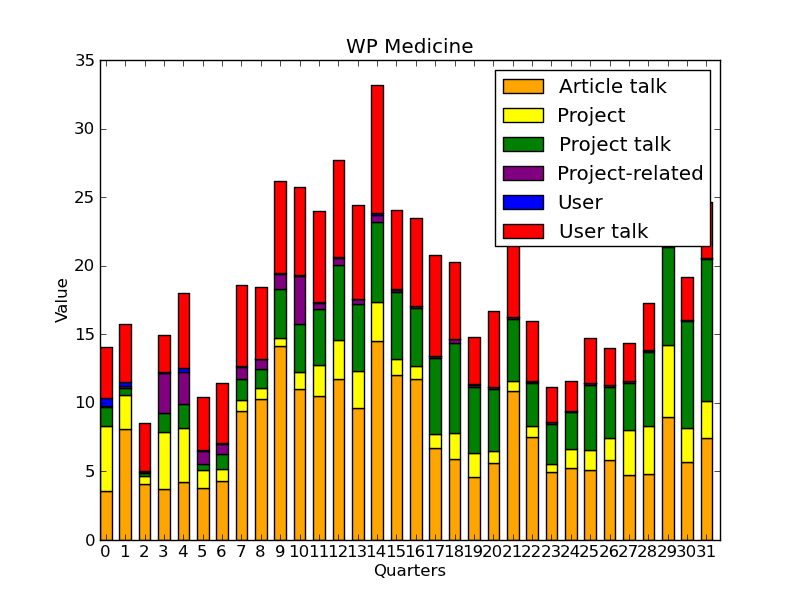}}\quad
  \subfloat{\includegraphics[width=0.90\textwidth]{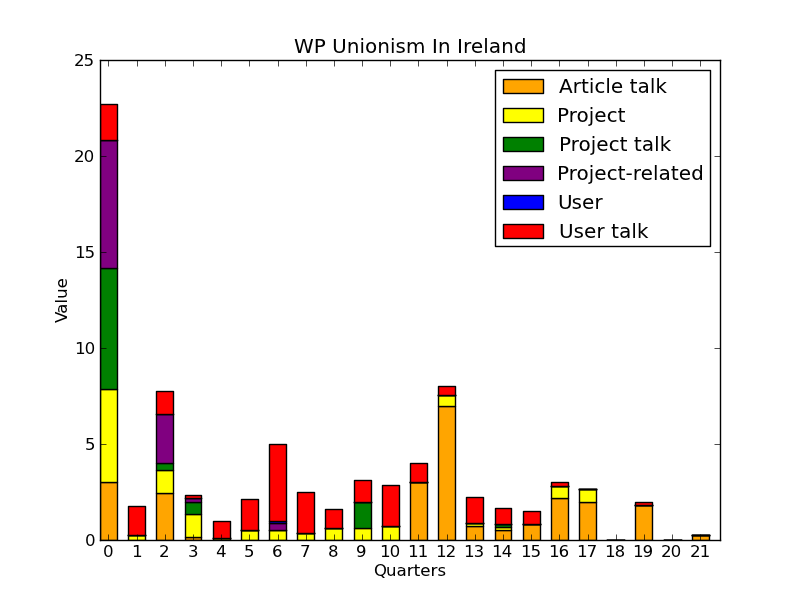}}\\
\caption{Evolution of Edit Distribution for WP Medicine and WP Unionism in Ireland.}
\label{fig:Edit_Distribution_Exploration}
\end{center}
}
\end{figure}
To further understand the characteristics of quarterly observations that locate at different areas (specifically, the far left, middle and top right area) of the curve in Figure \ref{fig:sc_losses_productivity}, we explore the dataset in more details by tracking the evolution of independent variables and average task distribution over quarters. Following other studies, we also quantify different types of tasks as edits to different types of pages within Wikipedia. The file system of Wikipedia is subdivided into ``namespaces"\footnote{http://en.wikipedia.org/wiki/Wikipedia:Namespace} which represent general categories of pages based on their function. Average task distribution is calculated as the average number of edits to a specific namespace by project members in a quarter. We observe that healthy WikiProjects generally experience lower levels of member turnover and social capital losses, thus enjoy better group outcomes and better task distribution, and tend to locate somewhere between the far left and middle areas of the curve in Figure \ref{fig:sc_losses_productivity}; while unhealthy WikiProjects generally experience more fluctuation in member turnover and social capital losses, have extremely unbalanced task distribution, and have their quarterly observations distributed over the three areas of the curve in Figure \ref{fig:sc_losses_productivity}.
\begin{figure}[!htb]
\baselineskip=12pt
{\fontsize{10.0pt}{10.0pt}\selectfont
\begin{center}
  \subfloat{\includegraphics[width=0.90\textwidth]{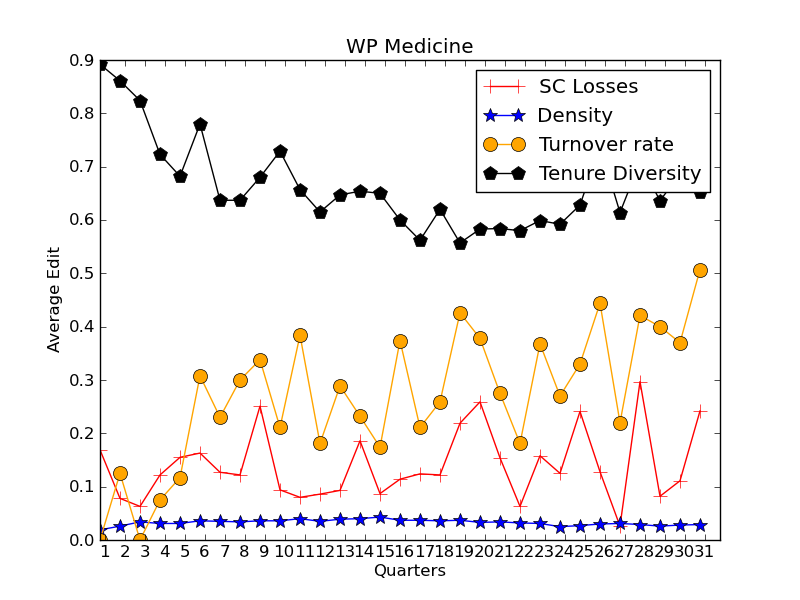}}\quad
  \subfloat{\includegraphics[width=0.90\textwidth]{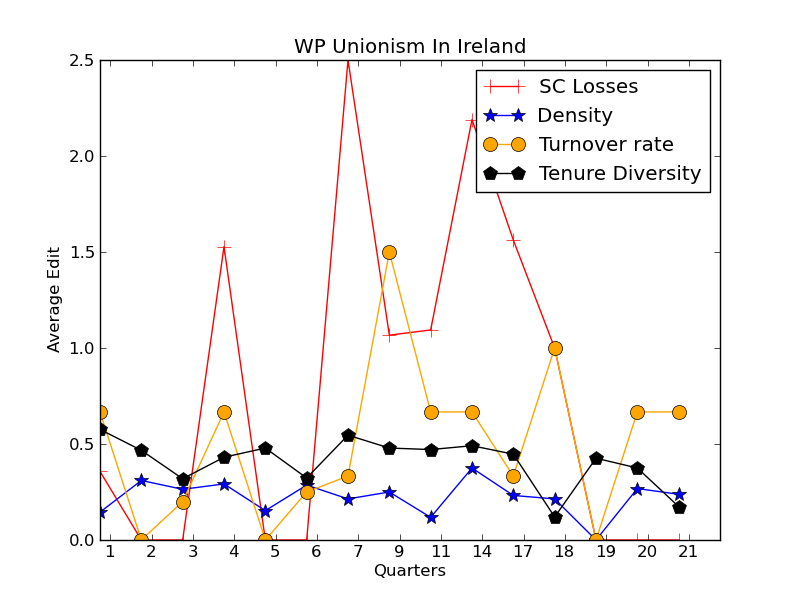}}\\
\caption{Evolution of Independent Variables for WP Medicine and WP Unionism in Ireland.}
\label{fig:Independent_Variable_Exploration}
\end{center}
}
\end{figure}

We found that WikiProject Medicine is a very healthy project with all its observations located at the far left region of the curve, while WikiProject Unionism in Ireland is a less healthy project. Figure \ref{fig:Edit_Distribution_Exploration} and Figure \ref{fig:Independent_Variable_Exploration} present the evolution of edit distribution and independent variables for the two projects, respectively. Because article edits generally account for more that 50\% of members' activity, we omitted it from Figure \ref{fig:Edit_Distribution_Exploration} to improve readability of the graph. The two figures in Figure \ref{fig:Edit_Distribution_Exploration} shows a substantial difference among their average edit distribution: WikiProject Medicine enjoys better and more balanced task distribution than WikiProject Unionism in Ireland, with the members of the former making edits across a range of namespaces each quarter. The figures in Figure \ref{fig:Independent_Variable_Exploration} reveals that WikiProject Unionism in Ireland experiences substantial fluctuation in its member turnover and social capital losses, while WikiProject Medicine has a more stable evolution in its independent variables.

\section{Conclusion}
We have empirically examined the impact of membership turnover in online groups on its productivity. The results show that membership turnover has significant negative effect on group outcomes in Wikipedia, and that social capital losses through member turnover are associated with group productivity in a negative curvilinear fashion. Our results show that the loss of social capital in collaboration networks can explain significant amount of variation in productivity for WikiProjects. The results contribute to the existing literature that investigates the relationship between social capital losses and group performance (e.g., \cite{Huselid1995,Abelson1984,Shaw2005,Ransbotham2011}) in conventional organizations and online communities. The results also support the mediation of turnover rate and network density on the relationship between social capital losses and group productivity.

Previous studies have suggested that editors with different tenure are likely to perform different types of tasks \citep{Bryant2005}, and that groups consisting of members with diversified tenure generally enjoy better task distribution and group performance \citep{Chen2010}. Similarly, our exploration of the dataset reveals that projects with low levels of social capital losses and turnover rate generally enjoy better task distribution and group outcomes. It will be very interesting to investigate the impact of task distribution among members on group performance and to study the evolution of user behaviours in online communities. Furthermore, we observe that project members denote substantial amount of time to edit non main namespaces such as article talk pages and user talk pages. This study only considers social capital losses in collaboration networks, it will be worthwhile to take into account social capital losses in communication networks for WikiProjects and in article talk networks and user talk networks.

\section{Acknowledgements}

This work is supported by Science Foundation Ireland (SFI) under Grant Number SFI/12/RC/2289 (Insight Centre). Xiangju Qin is funded by University College Dublin and China Scholarship Council (UCD-CSC Scholarship Scheme 2011).

\bibliographystyle{aaai_bib}
\bibliography{wikiproject_ref}

\end{document}